\documentclass[sigconf]{acmart}

\usepackage{booktabs}
\usepackage{multirow}
\usepackage{caption}
\usepackage{enumitem}
\usepackage{listings}
\usepackage{graphicx}
\usepackage{balance}
\usepackage{hyperref}
\usepackage{cleveref}
\usepackage{threeparttable}
\usepackage{tcolorbox}
\usepackage{xcolor}
\AtBeginDocument{%
  }
    
\setcopyright{none} 
\renewcommand\footnotetextcopyrightpermission[1]{}
\begin{document}

\title{The Hidden Cost of Readability: How Code Formatting Silently Consumes Your LLM Budget }

\author{Dangfeng Pan}
\authornote{Both authors contributed equally to this research.}
\orcid{0009-0004-1958-4314}
\affiliation{%
  \institution{Monash University}
  \country{Australia}
}
\email{dpan0026@student.monash.edu}

\author{Zhensu Sun}
\authornotemark[1]
\affiliation{%
  \institution{Singapore Management University}
  \country{Singapore}}
\email{zssun@smu.edu.sg}

\author{Cenyuan Zhang}
\affiliation{%
  \institution{Monash University}
  \country{Australia}
}
\email{cenyuan.zhang@monash.edu}

\author{David Lo}
\affiliation{%
  \institution{Singapore Management University}
  \country{Singapore}}
\email{davidlo@smu.edu.sg}

\author{Xiaoning Du}
\authornote{Corresponding author}
\affiliation{%
  \institution{Monash University}
  \country{Australia}
}
\email{xiaoning.du@monash.edu}

\begin{abstract}
Source code is usually formatted with elements like indentation and newlines to improve readability for human developers.
However, these visual aids do not seem to be beneficial for large language models (LLMs) in the same way since the code is processed as a linear sequence of tokens.
Furthermore, these additional tokens can lead to increased computational costs and longer response times for LLMs.
If such formatting elements are non-essential to LLMs, we can reduce such costs by removing them from the code.
To figure out the role played by formatting elements, we conduct a comprehensive empirical study to evaluate the impact of code formatting on LLM performance and efficiency.
Through large-scale experiments on Fill-in-the-Middle Code Completion tasks across four programming languages (Java, Python, C++, C\#) and ten LLMs—including both commercial and open-source models—we systematically analyze token count and performance when formatting elements are removed.
Key findings indicate that LLMs can maintain performance across formatted code and unformatted code, achieving an average input token reduction of 24.5\% with negligible output token reductions.
This makes code format removal a practical optimization strategy for improving LLM efficiency.
Further exploration reveals that both prompting and fine-tuning LLMs can lead to significant reductions (up to 36.1\%) in output code length without compromising correctness.
To facilitate practical applications, we develop a bidirectional code transformation tool for format processing, which can be seamlessly integrated into existing LLM inference workflows, ensuring both human readability and LLM efficiency.
\end{abstract}

\maketitle

\section{Introduction}

Large Language Models (LLMs) have revolutionized software development through their remarkable capabilities in code understanding and code generation. 
When tasked with code generation or completion, LLMs can interpret a developer's intent from incomplete code snippets or natural language descriptions, producing code suggestions that align closely with the developer's expectations~\cite{sun2024neural, sun2025don}.
Recently, advanced models such as GPT-4o and Gemini-1.5 have shown performance levels comparable to those of human programmers across a range of programming tasks and languages~\cite{advs.202412279}.
LLMs focus on the next-token-prediction task during pretraining, where they learn from vast and diverse textual corpora and become generalizable to various tasks and domains~\cite{qi2024tokenpredictionsufficientgpt}.
On the other hand, this paradigm imposes restrictions on the representation of content, particularly for information dimensions that cannot be fully or effectively captured in a linear fashion~\cite{opmi_a_00160}.

\begin{figure}
    \centering
    \includegraphics[width=1\linewidth]{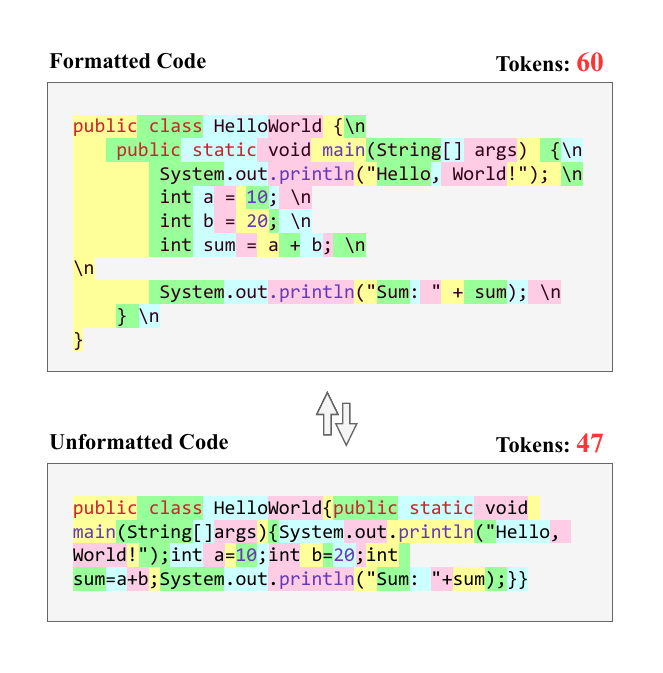}
    \vspace{-6mm}
    \caption{A comparison between formatted and unformatted Java code snippet, tokenized by GPT-4o's tokenizer.  Continuous characters with the same background color represent the same token. The unformatted code is produced by removing indentation, whitespaces, and newlines while maintaining syntactic correctness.}

    \label{fig:format-comparison}
\end{figure}

Code differs from natural language due to its structured nature. 
Since code can be quite complex, adhering to appropriate coding styles is crucial for ensuring readability. 
Specific conventions are usually developed for programming in different languages, defining coding rules that have been shown to be most beneficial for understanding and maintaining the code.
Recommended code styles visually enhance the identification of the code structure without affecting the code semantics, thereby aiding in the comprehension of its meaning~\cite{sun2024aicodersusrethinking}. 
Formatting elements such as indentation, whitespace, and newlines are commonly used to achieve this clarity. 
However, these formatting practices can increase the overall length of the code, leading to more tokens being processed by language model tokenizers. 
The visual advantages of well-formatted code can be lost in a linear representation of tokens, rendering the effort to use formatting somewhat ineffective. 
Additionally, the extra tokens consume a significant portion of the token budget in LLM-based code generation.

To have a quantitative understanding of the contribution of formatting elements in token count, we carried out a preliminary study.
We randomly sampled 100,000 source files for each programming language (Java, C\#, and C++) from the Stack v2~\cite{lozhkov2024starcoder2stackv2} code datasets, and measured their token counts using GPT-4o's tokenizer.
As a comparison, we also measure the token count when the three types of formatting elements - indentation, whitespace, and newline - are removed.
Notably, we only remove the ones that do not affect the code semantics, i.e., the AST before and after the removal remains the same.
We demonstrate the effect of format removal for a Java class definition in~\Cref{fig:format-comparison}.
We can observe that the token count is reduced from 60 to 47, indicating the formats incur 13 tokens' redundancy in the representation.
With the same measurement, we observe a surprisingly non-marginal token reduction for all three languages, with 14.7\% for Java, 13.2\% for C\#, and 13.2\% for C++.
In contrast, Python only shows a modest reduction of 4.0\%.
This limited reduction can be attributed to the fact that Python’s syntax relies heavily on indentation and newlines, which are required for syntactical correctness and thus cannot be removed.
Since LLMs operate on a token-by-token basis, the number of tokens to be processed or generated directly impacts their inference efficiency, where most commercial LLM APIs charge based on token count for both input and output \cite{AnthropicPricing2025, GoogleGeminiPricing2025}. 
Given this non-trivial overhead brought by formatting elements, it is critical to have a clear understanding of the role they play in code representations when being processed by LLMs.

To the best of our knowledge, there is still a limited understanding of this topic. 
Different studies have reported differing findings: some indicate that LLMs pay less attention to formatting elements~\cite{zeng2022extensive}, while others suggest that seemingly insignificant tokens may help LLMs make more informed decisions~\cite{goyal2023think}. 
In a recent paper~\cite{sun2024aicodersusrethinking}, the authors introduce a simplified AI-oriented Python grammar that removes both formatting elements and unnecessary grammar tokens.
However, it does not analyze the impact of formatting elements and requires further pre-training to adapt to the new grammar, leaving the effects of formatting elements on already-trained LLMs unclear.
To address this question more effectively, a large-scale, comprehensive experimental evaluation is necessary. 
Instead of relying on attention measurements, a more convincing approach would be to directly evaluate how model performance changes when formatting elements are either included or removed.

To fill this knowledge gap, we conduct a comprehensive empirical study to understand the role code formats play with respect to LLM's code generation capability. 
We employ a specific code completion task, the Fill-in-the-Middle (FIM), which shows the model a piece of incomplete code with missing middle sections and instructs it to generate the completion.
It is a common task for nearly all coding assistants in IDEs and can assess both the code understanding and generation capabilities of the LLM.
Our study will focus on four widely-used programming languages: Java, Python, C++, and C\#, all of which are commonly mastered by most LLMs. 
We identify the formatting elements that can be omitted for each language by examining their lexer configurations and grammar rules. 
To ensure that only relevant formatting elements in the actual code are retained for evaluation, we select elements categorized as non-essential or skippable, while excluding those that appear within comments or other non-code sections.
Finally, we target three formatting elements, i.e., indentation, whitespace, and newline.
It is important to note that our goal is to remove formatting elements that do not contribute any semantic meaning, ensuring that the code functionality conveyed by both the formatted and unformatted code remains unchanged.
The code before and after the removal is respectively named as \textbf{Formatted Code} and \textbf{Unformatted Code}.
The study includes ten models: five commercial API-based models and five open-weight models. 
We utilize McEval\cite{mceval}, a multilingual dataset for FIM tasks, which encompasses all four programming languages being studied and provides test cases for evaluation.
Next, we introduce the three RQs and summarize essential findings.

\smallskip

\noindent \textbf{RQ1: Can LLM maintain their performance when handling unformatted code, and how does code formatting impact their efficiency?}
In this RQ, we explore how well LLMs can maintain their performance when formatting elements are removed from the prompts of FIM tasks.
Additionally, we are interested in whether LLMs can adhere to the input code's formatting style when generating answers. 
If they can, it suggests that LLMs tend to maintain a consistent coding style during operations, which could be utilized to create more efficient LLMs.

\noindent \textit{Findings.} On average, across all settings, the performance of evaluated models remains largely unaffected by the removal of formatting. 
For example, DeepSeek-V3 shows minimal variation in Pass@1 scores across all programming languages, with an average of 79.1\% for formatted code and 80.0\% for unformatted code.
Regarding inference efficiency, removing formatting elements significantly reduces the number of input code tokens by 24.6\%, while the decrease in output code tokens is much smaller at only 2.9\%.
This indicates that the model tends to generate code in a familiar formatting style, regardless of that in the input.
In summary, removing formatting from the input has a minimal impact on the performance of LLMs and improves inference efficiency.
However, the efficiency gain could be enhanced if the models become more adaptable to different styles.

\smallskip
\noindent \textbf{RQ2: What is the impact of each formatting element on model performance and efficiency?}
Although removing all formatting elements has minimal performance impact, further ablation studies are necessary to determine if significant variance exists between the impact of different formatting elements.
In this RQ, through an ablation study, we assess the influence of removing one type of formatting element at a time on token consumption and model performance.

\noindent \textit{Findings.} Unlike the removal of all formatting elements, removing individual formatting elements can introduce negative impacts for some LLMs.
Specifically, Claude-3.7 and GPT-4 exhibit strong robustness, with performance variations (less than 1.6\%) remaining minimal, consistent with the findings when all formatting elements are removed.
In contrast, Gemini-1.5 shows significant performance drops when any single formatting element is removed, highlighting its sensitivity to partially formatted styles.
Additionally, while removing single formatting elements can effectively reduce input tokens, the reduction rate of output tokens remains low, similar to the case of completely unformatted code.
On average, output tokens decrease by only 0.4\% for Claude-3.7, 3.5\% for Gemini-1.5, and 1.4\% for GPT-4 when individual formatting elements are removed.
These limited reductions in output tokens, coupled with the performance degradation observed in Gemini-1.5, underscore the importance of adapting LLMs to different formatting styles.

\smallskip
\noindent \textbf{RQ3: How to enable LLMs to minimize token usage when generating outputs?}
In RQ1 and RQ2, we demonstrated that LLMs benefit from unformatted code input.
However, these models still prefer generating formatted code as output, regardless of the input format, leading to unnecessary token usage for formatting.
In RQ3, we explore adapting models for producing token-efficient code in the output, focusing on two cost-effective approaches: training-free prompting and fine-tuning (on very few samples).

\noindent \textit{Findings.} 
Prompting LLMs with clear instructions to generate unformatted output code can effectively reduce token usage while maintaining performance.
For instance, with a well-crafted prompt, GPT-4o achieves a significant reduction in output tokens (an average of 27.2\%) while maintaining performance in Java, C++, and C\#.
However, prompting can fail if the instructions are ambiguous or misinterpreted by the LLM.
For example, Gemini-1.5 often removes elements in a way that violates syntax rules, leading to a sharp decline in Pass@1 (e.g., from 67.2\% to 11.1\% in C++).
Similarly, fine-tuning with unformatted samples can also reduce output tokens while preserving or even improving Pass@1.
By fine-tuning with just 50 unformatted Java code samples, Gemini-1.5 and GPT-4o achieve substantial reductions in output tokens (35.9\% and 24.8\%, respectively) with statistically insignificant performance impact.
Both methods are feasible, and the choice between prompt engineering and fine-tuning depends on the LLM’s use case and the user’s role.

\smallskip

\begin{figure*}[t]
    \centering
    \includegraphics[width=0.99\linewidth]{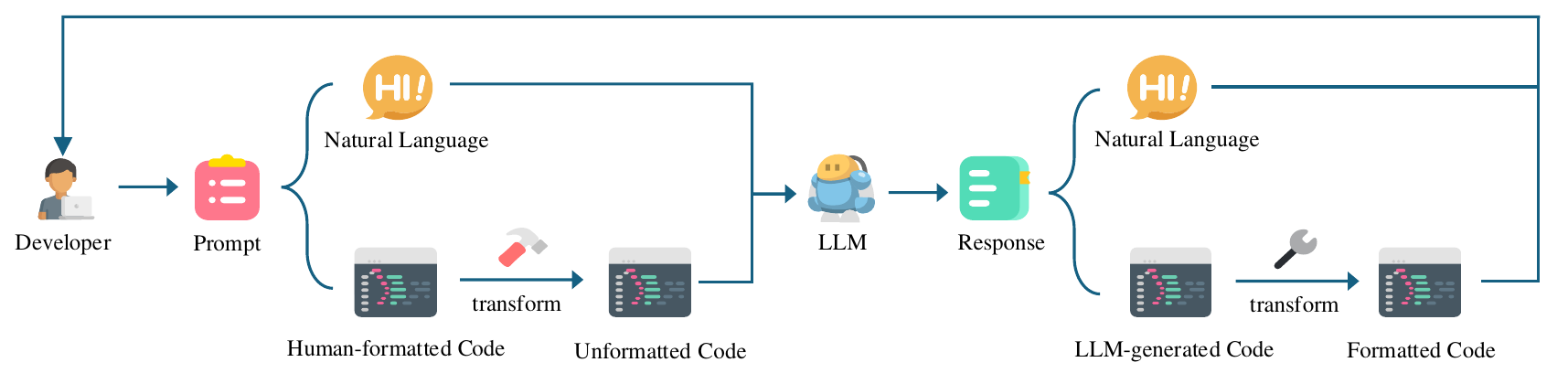}
    \caption{A demonstration of how unformatted code can fit in the existing Human-AI workflow.}
    \vspace{-4mm}
    \label{fig:tool_architecture}
\end{figure*}

These empirical results strongly suggest that code format can be and should be removed when LLMs work with source code.
To this end, we propose a code transformation tool that enables bidirectional conversion between formatted and unformatted code, preserving program semantics while reducing token overhead.
As shown in~\Cref{fig:tool_architecture}, this tool can transform human-formatted code into an unformatted, token-efficient representation for LLM processing, and then reformats the LLM-generated output into a human-readable format.
It enables developers to work with familiar, well-formatted code while LLMs work with token-efficient, unformatted code.
The tool currently supports four programming languages and has been well tested through rigorous AST equivalence verification across the entire McEval dataset.
It demonstrates an average transformation speed of 76ms per code sample, ensuring both semantic preservation and efficient real-time processing.

The tool code and live demo are available at~\url{https://sites.google.com/view/the-hidden-cost-of-readability}. Our work makes several key contributions:
\begin{itemize}[leftmargin=*]
    \item We reveal that using unformatted code can significantly reduce token usage without compromising the performance of LLMs, offering a practical optimization strategy for efficient LLM serving.
    \item  We demonstrate that LLMs can be trained or instructed to produce token-efficient code, further enhancing their efficiency in code generation tasks.
    \item We propose and implement a code transformation tool that facilitates bidirectional conversion between formatted and unformatted code, allowing developers to work with human-readable code while enabling LLMs to process code in a token-efficient manner.
\end{itemize}

\section{Experimental Setting}
In this section, we introduce the experimental setting for this study, including the task, benchmark, evaluation metrics, LLMs, format processing methods, and implementation details.
Our study is guided by three research questions:
\begin{itemize}[leftmargin=*]
    \item \textbf{RQ1:} Can LLMs maintain their performance when handling unformatted code, and how does code formatting impact their efficiency?
    \item \textbf{RQ2:} What is the impact of each formatting element on mode performance and efficiency?
    \item \textbf{RQ3:} How to enable LLMs to minimize token usage when generating outputs?
\end{itemize}

\subsection{Task and Benchmark}
We choose the \textbf{Fill-in-the-Middle} (FIM) code completion task~\cite{bavarian2022efficienttraininglanguagemodels}, the most prevalent and practical code completion paradigm used by commercial AI-powered programming assistants~\cite{GitHub2023}, to investigate the impact of code format.
The task requires the LLM to complete the code snippets with missing middle sections, which can assess the model's code understanding and generation capability at the same time.
Given the significant costs associated with using commercial LLM APIs, our experiment includes four programming languages: Java, Python, C++, and C\#, which feature diverse formatting conventions and are among the most popular languages in production environments.
Catering to these needs, we select \textbf{McEval}\cite{mceval} as the benchmark for our study.
It is created by professional developers through a rigorous annotation and verification process, covering 40 programming languages.
By having experienced developers manually create and validate each sample, the dataset is independently generated rather than being automatically translated from Python, which enhances diversity and authenticity while minimizing the risk of data leakage from training sets. 
In this benchmark, each sample consists of a code snippet with a missing middle section for the FIM task, a detailed problem description that explains the coding challenge, and a set of corresponding test cases to evaluate the correctness of the generated code. 
We use its subset for the four languages, which offers 314 samples for C++, 318 for C\#, 355 for Java, and 330 for Python.

\subsection{Large Language Models}

To ensure the generalization of our findings, we evaluate a diverse range of state-of-the-art LLMs, including five commercial API-based models and five open-weight models. 
Among the commercial models, we include three models from OpenAI: \textbf{GPT-3.5-turbo}, a widely used efficient model that balances performance and cost-effectiveness in production applications; \textbf{GPT-4o-mini}, a mid-sized model that strikes a balance between computational efficiency and robust capabilities; and \textbf{GPT-4o}, an advanced multimodal model representing the state-of-the-art in commercial LLM performance \cite{openai2024gpt4ocard,openai2024gpt4technicalreport}. 
Additionally, we evaluate Google’s \textbf{Gemini-1.5}, specifically using the gemini-1.5-flash version, which is a fast and versatile multimodal model for scaling across diverse tasks. \cite{geminiteam2024gemini15unlockingmultimodal}, and Anthropic’s \textbf{Claude-3.7}, which emphasizes safety and alignment \cite{Claude3}.

On the open-weight front, the five models are \textbf{Phi-3.5} (with 3.82B parameters), a lightweight instruction-tuned model pre-trained on synthetic data and filtered web content for reasoning and instruction-following tasks \cite{arxivid:2404.14219}; 
\textbf{Qwen-2.5} (1.54B parameters), Alibaba’s specialized instruction-tuned model optimized for code generation and reasoning \cite{qwen2};
\textbf{MagiCoder}, an instruction-tuned system (6.7B parameters) fine-tuned from Deepseek-coder-6.7B using the OSS-Instruct method to enhance code generation quality \cite{arxivid:2312.02120};
and two models from DeepSeek: \textbf{DeepSeek-V3}, a powerful Mixture-of-Experts model with 671B total parameters (37B activated per token) pre-trained on 14.8 trillion tokens and fine-tuned via supervised and reinforcement learning \cite{deepseekai2025deepseekv3technicalreport}, and \textbf{Deepseek-coder-1.3B}, trained from scratch on 2 trillion tokens with a composition of 87\% code and 13\% natural language \cite{guo2024deepseekcoderlargelanguagemodel}.

\subsection{Evaluation Metrics}\label{sec:metric}
We evaluate our approach using the following metrics:

\begin{itemize}[leftmargin=*]
\item \textbf{Pass@1:} To compute Pass@1, one code sample is generated for each problem in the benchmark, and a problem is considered solved if the samples pass the unit tests. 
We report the fraction of problems being successfully solved.
        
\item \textbf{Token Counts:} We assess efficiency by counting the tokens in the input code and the generated output. 
Notably, since tokens are processed using each model's tokenizer, the same text may yield different token counts across different LLMs.
For input token reduction, we extract the code portion from the prompt and calculate the difference in token counts between the formatted and unformatted versions.
For output token reduction, directly comparing outputs generated from formatted and unformatted inputs may introduce bias to our conclusions, since any differences in the input (e.g., due to formatting) can shift the model’s attention and alter its generation trajectory.
To avoid this, we estimate output token reduction by reformatting the LLM-generated code for unformatted input into the most concisely formatted version and comparing the token counts before and after reformatting.
This approach quantifies the potential token savings in the output.
Notably, token counts can be directly translated to the financial cost when using commercial LLM APIs.
For example, GPT-4o charges \$2.5 per 1M tokens for input and \$10.00 per 1M tokens for output~\cite{opanaiapipricing}.

\item \textbf{Statistical Validation:} We employ McNemar's test~\cite{McNemar1947}, a non-parametric method particularly suited for comparing the performance of two models on the same test instances.
This test helps us determine whether differences in Pass@1 between the experimental and control groups are statistically significant ($p$-value < $0.05$).
Additionally, for analyzing the statistical significance of differences in input and output token counts between these experimental and control groups, we utilize the Mann-Whitney U Test~\cite{mann1947test}.
This non-parametric test is appropriate for comparing distributions of token counts as it does not assume normality in the data, allowing us to robustly assess whether the observed token reductions are statistically significant.
To address the multiple testing problem across different RQs and measurements, we apply False Discovery Rate (FDR) correction using the Benjamini-Hochberg procedure~\cite{benjamini1995controlling}, which controls the expected proportion of false positives and strengthens the validity of our statistical claims.
\end{itemize}

\subsection{Format Processing}
We clarify the formatting elements considered in our study and clearly define formatted code and unformatted code.

\begin{itemize}[leftmargin=*]
\item \textbf{Formatting elements.}
To identify omittable formatting elements for evaluation, we analyze the lexer configurations and grammar rules of each programming language, selecting elements categorized as non-essential or skippable.
After excluding elements related to comments or other non-code components, three formatting elements are selected for evaluation: indentation, whitespace, and newlines.
In Java, C\#, and C++, indentation, newlines, and additional whitespace are formatting tokens that do not affect the semantic meaning of the code, except in the case of preprocessor directives in C\# and C++. For Python, only whitespace is removed, as the other two elements are required by Python syntax.

\item \textbf{Formatted Code (Control Group):}
To establish uniform control groups, we standardize the usage of the formatting elements mentioned above in all code samples according to the Google Style guidelines~\cite{GoogleStyleGuide} for C-family languages (C++, C\#, Java) and PEP 8~\cite{PEP8} for Python. This ensures that our formatting accurately reflects real-world practices.
This standardization results in a slight increase of 1.15\% in the token count of the evaluation datasets. 
While this may affect the quantitative results of our experiments, it allows for more informed and transparent conclusions. 
We believe that the qualitative findings remain valid despite these slight changes in formatting rules.

\item \textbf{Unformatted Code (Experimental Group):}
For all code samples, we minimize whitespaces, newlines, and indentation without violating the syntax rules of each programming language.
Regarding whitespaces, we eliminate non-essential spaces while keeping those required by the language's syntax (e.g., spaces between keywords and identifiers). 
For newlines, we remove blank lines and combine multiple statements onto single lines wherever syntactically permissible. We preserve necessary newlines, such as those separating preprocessor directives in C++.
As for indentation, we remove all leading spaces or tabs at the beginning of lines, as long as the syntax allows.

\end{itemize}

\subsection{Implementation Details}
In our experiments, we evaluate a diverse range of popular LLMs, including commercial API-based models and open-weight models.
The commercial API-based models include GPT-3.5-turbo, GPT-4o-mini, GPT-4o from OpenAI, Gemini-1.5 from Google, Claude-3.7 from Anthropic, and DeepSeek-V3 from Novita (a third-party API provider).
For API invocation, we implement standardized REST requests to each provider's endpoints using their official SDKs with default parameters. 
Additionally, we evaluate open-weight models locally, including Phi-3.5, Qwen-2.5, Deepseek-coder, and Magicoder.
The local models are implemented using the Huggingface Transformers library with PyTorch and executed on hardware equipped with 40 vCPUs, 480GB RAM, and an NVIDIA GeForce RTX 3090 GPU (24GB VRAM).
For these local models, the maximum token limit is set to 2048 due to GPU memory constraints.
During inference, we use default hyperparameters for all models, setting the temperature to 0 to ensure deterministic outputs.
For fine-tuning experiments, we randomly select 50 examples from the McEval benchmark for each language (these selected examples are thus excluded in the corresponding evaluation) and fine-tune the commercial models through their fine-tuning APIs.

\section{Study Results}
In this section, we report our experimental results and answer the three research questions.

\subsection{RQ1: Impact of Unformatted Code on LLM Performance and Efficiency}
In this RQ, we systematically evaluate the impact of removing formatting elements from the prompts of FIM tasks.
Specifically, we evaluate all the ten models using the McEval benchmark of the four programming languages, C++, C\#, Java, and Python.
We prompt the models with two versions of the incomplete code snippet, namely, the formatted version and the unformatted version.
We calculate Pass@1 using test cases and measure token counts for both the incomplete code and the generated completions, following the methodology outlined in~\Cref{sec:metric}.
Additionally, we assess the statistical significance between the Pass@1 scores and token counts of the experimental group (using unformatted code) and the control group (using formatted code).

The results are presented in \Cref{tab:formatting_impact}.
We bold case where the performance on unformatted input code is either equal to or only marginally lower (within a 2\% margin) than the performance on formatted code.

\begin{table*}[ht]
\centering
\setlength{\tabcolsep}{4pt}
\caption{Comparison of model performance (Pass@1) and token efficiency between formatted code (F) and unformatted code (U). Models include DeepSeek-V3 (DS-V3), Claude-3.7, Gemini-1.5, MagiCoder (MC), GPT-4o, GPT-4o-mini (GPT-4o-m), DeepSeek-Coder (DS-C), Qwen-2.5, GPT-3.5, and Phi-3.5.}

\begin{tabular}{llllllllllll}
\toprule
\textbf{Language} & \textbf{Metric} & \textbf{DS-V3} & \textbf{Claude} & \textbf{Gemini} & \textbf{MC} & \textbf{GPT-4o} & \textbf{GPT-4o-m} & \textbf{DS-C} & \textbf{Qwen} & \textbf{GPT}\textbf{-3.5} & \textbf{Phi} \\
\midrule
\multirow{4}{*}{\textbf{C++}} 
& Pass@1 (F) & \textbf{76.1\%} & \textbf{72.9\%} & \textbf{68.5\%} & \textbf{65.9\%} & \textbf{63.7\%} & \textbf{62.4\%} & 49.0\% & \textbf{39.2\%}$^{*}$ & \textbf{31.5\%}$^{*}$ & 16.9\% \\
& Pass@1 (U) & \textbf{76.8\%} & \textbf{72.9\%} & \textbf{67.2\%} & \textbf{64.3\%} & \textbf{61.8\%} & \textbf{62.4\%} & 46.8\% & \textbf{46.8\%}$^{*}$ & \textbf{38.2\%}$^{*}$ & 12.7\% \\
& Input Reduction & 30.8\%$^{**}$ & 28.9\%$^{**}$ & 34.0\%$^{**}$& 24.8\%$^{**}$ & 33.7\%$^{**}$ & 33.7\%$^{**}$ & 24.8\%$^{**}$ & 34.4\%$^{**}$ & 35.1\%$^{**}$ & 31.0\%$^{**}$ \\
& Output Reduction & 0.0\% & -0.7\% & 9.3\% & 2.9\% & 0.4\% & 0.7\% & 2.6\% & 1.9\% & 2.8\% & 1.3\% \\
\midrule
\multirow{4}{*}{\textbf{C\#}} 
& Pass@1 (F) & \textbf{87.7\%} & 90.3\% & \textbf{77.4\%} & \textbf{73.6\%} & \textbf{76.7\%} & \textbf{78.3\%} & 51.6\% & \textbf{45.6\%} & \textbf{47.8\%} & \textbf{15.4\%} \\
& Pass@1 (U) & \textbf{86.8\%} & 87.7\% & \textbf{76.4\%} & \textbf{73.6\%} & \textbf{77.7\%} & \textbf{77.7\%} & 48.7\% & \textbf{49.1\%} & \textbf{50.6\%} & \textbf{15.7\%} \\
& Input Reduction & 22.3\%$^{**}$ & 22.7\%$^{**}$ & 29.7\%$^{**}$& 22.7\%$^{**}$ & 26.2\%$^{**}$ & 26.2\%$^{**}$ & 22.7\%$^{**}$ & 26.8\%$^{**}$ & 27.1\%$^{**}$ & 26.2\%$^{**}$ \\
& Output Reduction & -0.5\% & -2.2\% & 2.7\% & 0.7\% & -0.5\% & -0.4\% & 0.7\% & 1.2\% & 0.5\% & 0.4\% \\
\midrule
\multirow{4}{*}{\textbf{Java}} 
& Pass@1 (F) & \textbf{70.7\%} & \textbf{68.5\%} & \textbf{67.9\%} & \textbf{63.7\%} & \textbf{63.1\%} & \textbf{65.1\%} & \textbf{56.1\%} & \textbf{47.9\%} & \textbf{36.9\%}$^{*}$ & \textbf{15.8\%} \\
& Pass@1 (U) & \textbf{71.3\%} & \textbf{69.0\%} & \textbf{67.9\%} & \textbf{63.7\%} & \textbf{65.6\%} & \textbf{66.5\%} & \textbf{58.9\%} & \textbf{51.3\%} & \textbf{43.4\%}$^{*}$ & \textbf{17.7\%} \\
& Input Reduction & 29.9\%$^{**}$ & 35.7\%$^{**}$ & 42.0\%$^{**}$& 33.1\%$^{**}$ & 33.9\%$^{**}$ & 33.9\%$^{**}$ & 33.1\%$^{**}$ & 35.0\%$^{**}$ & 35.1\%$^{**}$ & 37.2\%$^{**}$ \\
& Output Reduction & 3.7\% & 5.0\% & 11.5\% & 6.8\% & 4.6\% & 4.1\% & 7.2\% & 7.3\% & 4.6\% & 6.2\% \\
\midrule
\multirow{4}{*}{\textbf{Python}} 
& Pass@1 (F) & \textbf{81.8\%} & \textbf{84.5\%} & \textbf{71.2\%} & \textbf{44.2\%} & \textbf{66.4\%}$^{*}$ & \textbf{52.1\%} & \textbf{50.9\%} & 66.7\% & 73.6\% & \textbf{20.9\%} \\
& Pass@1 (U) & \textbf{85.2\%} & \textbf{87.0\%} & \textbf{71.8\%} & \textbf{44.8\%} & \textbf{71.5\%}$^{*}$ & \textbf{51.8\%} & \textbf{49.7\%} & 63.9\% & 70.3\% & \textbf{20.3\%} \\
& Input Reduction & 7.4\% & 4.7\% & 5.6\% & 2.2\% & 9.4\% & 9.4\% & 2.2\% & 9.5\% & 9.5\% & 5.2\% \\
& Output Reduction & 0.0\% & -1.1\% & 4.1\% & 1.9\% & 0.4\% & 0.5\% & 1.4\% & 4.3\% & 2.3\% & 0.1\% \\
\bottomrule
\end{tabular}

{\footnotesize \makebox[\textwidth][r]{$Bold$: Unformatted code performance is either better than or within 2\% of formatted code.\quad\quad$^{*}$ : $p$-value $< 0.05\quad\quad$ $^{**}$ : $p$-value $< 0.01\quad\quad$}}

\label{tab:formatting_impact}
\end{table*}

\subsubsection{Impacts on LLM performance}
Most models exhibit stable performance across formatted and unformatted code inputs, with only minor fluctuations observed in certain languages. 
Specifically, across all models and languages, we do not observe a Pass@1 drop higher than 4.2\% after removing the code format in the input, and all those drops exhibit insignificant $p$-values.
State-of-the-art (SOTA) models, such as Deepseek-V3, Gemini-1.5, and GPT-4o, demonstrate negligible differences in Pass@1 scores between formatted and unformatted code, indicating that removing formatting has no significant impact on their performance.
Smaller or less performant models exhibit more pronounced fluctuations. 
For example, Deepseek-Coder shows performance degradation in C++ (3.2\% drop in Pass@1) and C\# (2.9\% drop), while Phi-3.5 struggles with C++ (4.2\% drop).
This may be caused by the inherent limitations of smaller models in generalizing across diverse input structures.
In contrast, SOTA models, with their larger parameter counts and more advanced training methodologies, are better equipped to handle unformatted inputs.
Interestingly, we also observe cases where performance improves after removing the format, some of which are even statistically significant.
For instance, in Python, GPT-4o's performance increases from 66.4\% to 71.5\% when transitioning from formatted to unformatted code.
We infer that the removal of formatting elements simplifies the input in certain cases, reducing distractions from non-essential tokens and allowing models to focus on the core logic.
Investigating the reasons behind this phenomenon could be an interesting direction for future work.

\begin{tcolorbox}[size=title]
    {\textbf{Takeaway\#1:}}
    The code format in the input does not negatively impact the performance of LLMs, as they demonstrate comparable Pass@1 scores across both formatted and unformatted inputs.
\end{tcolorbox}

Among the evaluated languages, Java stands out as the most stable in terms of model performance when transitioning from formatted to unformatted code inputs.
This stability is evident across all models, with no significant performance drops observed even for smaller or less performant models.
This exceptional stability can likely be attributed to the prevalence of unformatted Java code in their pre-training datasets.
Java is one of the most widely used programming languages, and its code is frequently shared in diverse formats in open-source repositories, forums, and documentation.
As a result, models are likely exposed to a significant amount of unformatted or minimally formatted Java code during training, enabling them to better generalize such inputs.
In contrast, other languages exhibit greater sensitivity to format changes.
For example, both Qwen-2.5 and GPT-3.5 show noticeable performance drops in Python (2.8\% and 3.3\% decrease in Pass@1, respectively), highlighting the influence of language-specific syntax and training data distribution on model performance.

\begin{tcolorbox}[size=title]
    {\textbf{Takeaway\#2:}}
    Models exhibit language-specific sensitivity to code formatting changes, where Java demonstrates exceptional stability across formatted and unformatted inputs.
\end{tcolorbox}

\subsubsection{Impacts on Token Reduction}
Using unformatted code as input can reduce a considerable number of tokens across all evaluated models, where the extent of this reduction varies due to differences in tokenizers. 
On average, the input code tokens are reduced by 25.8\% across all settings.
For Deepseek-V3, the input code token reduction is particularly notable, with Java code tokens reduced by 42.0\%, highlighting the potential of format removal in accelerating the models' code understanding speed.
Similarly, commercial LLMs like Claude-3.7, Gemini-1.5, and GPT-4o also demonstrate substantial input token reduction, with averages of 23.0\%, 27.8\%, and 25.8\%, respectively.
For these commercial models, these reductions directly translate to cost savings in their API services.
Take Claude-3.7 Sonnet for example.
It charges \$3 per 1M input tokens, where a 23.0\% token reduction can save \$0.69 per 1M tokens~\cite{AnthropicPricing2025}.
This cost efficiency makes format removal an attractive option for users of these models.
Moreover, the efficiency gains vary substantially depending on the programming language.
Java shows the highest efficiency gains, averaging 34.9\% token reduction, due to its verbose syntax and reliance on formatting for readability.
The efficiency gains are also substantial for C++ and C\#, respectively 31.12\% and 25.26\%.
In contrast, Python achieves only a 6.51\% average code token reduction, as formatting elements such as newlines and indentations are integral to its grammar and cannot be removed without compromising code functionality.
This language-specific variability must be carefully considered when deciding whether to use unformatted code for specific LLM-driven applications.

\begin{tcolorbox}[size=title]
    {\textbf{Takeaway\#3:}}
Removing code formatting can substantially reduce input tokens for languages that do not deeply integrate formatting elements into their syntax (e.g., Java, C++, C\#) and only slightly reduce tokens for languages like Python, where formatting is essential to syntax and functionality.
\end{tcolorbox}

\begin{table*}[t]
\centering
\caption{Output Token Reduction (\%) Across Languages and Models (All, Pass, Fail)}
\begin{tabular}{llcccccccccc}
\toprule
\textbf{Language} & \textbf{Condition} & \textbf{DS-V3} & \textbf{Claude} & \textbf{Gemini} & \textbf{MC} & \textbf{GPT-4o} & \textbf{GPT-4o-m} & \textbf{DS-C} & \textbf{Qwen} & \textbf{GPT-3.5} & \textbf{Phi} \\
\midrule
\multirow{3}{*}{\textbf{C++}} 
& Pass & -0.2\% & -0.6\% & 8.0\% & 3.1\% & -0.1\% & -0.1\% & 2.6\% & 2.6\% & 1.2\% & 1.3\% \\
& Fail & 0.5\% & -0.9\% & 11.3\%& 2.6\% & 1.4\% & 1.5\% & 2.6\% & 1.3\% & 3.9\% & 1.3\% \\
& All  & 0.0\% & -0.7\% & 9.3\% & 2.9\% & 0.4\% & 0.7\% & 2.6\% & 1.9\% & 2.8\% & 1.3\% \\
\midrule
\multirow{3}{*}{\textbf{C\#}} 
& Pass &  -0.5\% & -2.3\% & 3.0\% & 0.8\% &  -0.5\% &  -0.4\% & 0.4\% & 1.7\% & 0.1\% & 2.1\% \\
& Fail &  -0.6\% & -2.2\% & 1.7\% & 0.5\% &  -0.4\% &  -0.1\% &  1.0\% & 0.8\% & 1.0\% &  0.0\% \\
& All  &  -0.5\% & -2.2\% & 2.7\% & 0.7\% &  -0.5\% &  -0.4\% & 0.7\% & 1.2\% & 0.5\% & 0.4\%  \\
\midrule
\multirow{3}{*}{\textbf{Java}} 
& Pass & 3.2\% & 4.7\% & 11.2\%& 6.0\% & 4.0\% & 3.5\% & 7.0\% & 7.2\% & 4.6\% & 6.3\% \\
& Fail & 4.5\% & 5.6\% & 11.9\%& 7.7\% & 5.4\% & 5.0\% & 7.5\% & 7.4\% & 4.6\% & 6.2\% \\
& All  & 3.7\% & 5.1\% & 11.5\%& 6.8\% & 4.6\% & 4.1\% & 7.2\% & 7.3\% & 4.6\% & 6.2\%  \\
\midrule
\multirow{3}{*}{\textbf{Python}} 
& Pass & 0.1\% &  -1.1\% & 4.2\% & 1.7\% & 0.4\% & 0.6\% & 1.3\% & 4.0\% & 2.5\% & 0.1\% \\
& Fail & 0.1\% &  -1.1\% & 3.8\% & 2.0\% & 0.4\% & 0.3\% & 1.4\% & 4.6\% & 1.9\% & 0.1\% \\
& All  & 0.1\% &  -1.1\% & 4.1\% & 1.8\% & 0.4\% & 0.5\% & 1.4\% & 4.3\% & 2.3\% & 0.1\%  \\

\bottomrule
\end{tabular}

{\footnotesize \makebox[\textwidth][r]{\quad\quad$^{*}$ : $p$-value $< 0.05\quad\quad$ $^{**}$ : $p$-value $< 0.01\quad\quad$}}
\vspace{-4mm}
\label{tab:rq1_output_analysis}
\end{table*}

While input token reduction is substantial, output token reduction remains modest, depending on the language or models.
Specifically, output code token reduction averages 2.5\% across all models and languages, with the highest reduction being 11.5\%, achieved by Gemini-1.5 in Java.
This suggests that models have internalized formatting conventions and maintain them in outputs regardless of input formatting.
It underscores the need for further optimization in output generation to fully capitalize on the efficiency gains achieved through input token reduction.
We also observe certain instances of negative reduction. 
Upon examining the generated code, we find that this is because the code generated by the models sometimes misaligns with the style guidelines we use.
As a result, when we convert the generated code into the style guideline format, additional tokens are introduced, leading to negative reductions in some cases.

Given that the Pass@1 for some settings is relatively low, e.g., 31.5\% for GPT-3.5 in C++, a potential concern arises: the modest reduction in output tokens might stem from low-quality code generated during the process.
LLMs may produce shorter responses when the code is hard to complete.
To investigate the effects of code quality on the output reduction, we further analyze the output token reduction grouped by whether the output passed functional correctness tests.
As shown in~\Cref{tab:rq1_output_analysis}, the average output reduction rates for Passing, Fail, and All are slightly different (2.3\%, 2.7\%, 2.5\%, respectively), but the conclusion remains the same, i.e., output reductions are significantly lower than input reduction.
These differences are not statistically significant, indicating that functional correctness has no systematic impact on formatting preservation in the output.

\begin{tcolorbox}[size=title]
    {\textbf{Takeaway\#4:}}
    LLMs tend to maintain formatting conventions in their outputs, regardless of input formatting.
\end{tcolorbox}

\subsection{RQ2: Contribution of Individual Formatting Elements}
\begin{table*}[t]
\centering
\caption{The Pass@1 and token reduction of three LLMs with various formatting configurations.}
\begin{tabular}{lllllllllll}
\toprule
\multirow{3}{*}[-1ex]{\textbf{Model}} & \multirow{3}{*}[-1ex]{\textbf{Format}} & \multicolumn{3}{c}{\textbf{Pass@1}} & \multicolumn{6}{c}{\textbf{Token Reduction}} \\ 
\cmidrule(lr){3-5} \cmidrule(lr){6-11}
 &  & \multirow{2}{*}[-0.5ex]{\textbf{C++}} & \multirow{2}{*}[-0.5ex]{\textbf{C\#}} & \multirow{2}{*}[-0.5ex]{\textbf{Java}} & \multicolumn{2}{c}{\textbf{C++}} & \multicolumn{2}{c}{\textbf{C\#}} & \multicolumn{2}{c}{\textbf{Java}} \\ 
\cmidrule(lr){6-11} 
 &  &  &  &  & \textbf{Input} & \textbf{Output} & \textbf{Input} & \textbf{Output} & \textbf{Input} & \textbf{Output} \\ 
\midrule
\multirow{4}{*}{Claude} 
  & Whitespaces Removed & 72.3\% & 87.7\% & 69.0\% & 5.2\%  & 0.4\%  & 3.2\%  & -2.4\%  & 3.3\%  & 4.3\%  \\ 
  & Indentations Removed & 72.2\% & 88.7\% & 68.2\% & 6.7\%  & -1.4\%  & 6.4\%  & -1.6\%  & 10.5\% & 4.4\%  \\ 
  & Newlines Removed & 74.5\% & 89.0\% & 69.0\% & 13.4\%$^{*}$ & -2.6\%  & 11.7\%$^{*}$ & -1.4\%  & 18.7\%$^{**}$ & 4.3\%  \\ 
  & \textbf{All Removed} & 72.9\% & 87.7\% & 69.0\% & 28.9\% & -0.7\%  & 22.7\% & -2.3\%  & 35.7\% & 5.1\%  \\ 
\midrule
\multirow{4}{*}{Gemini} 
  & Whitespaces Removed & 64.6\% & 71.3\%$^{*}$ & 63.7\%$^{*}$ & 6.3\%  & 5.4\%  & 3.4\%  & 2.2\%  & 3.4\%  & 6.0\%  \\ 
  & Indentations Removed & 66.6\% & 71.1\%$^{*}$ & 65.1\% & 7.3\%  & 2.5\%  & 7.5\%  & 0.5\%  & 11.8\% & 6.8\%  \\ 
  & Newlines Removed & 66.9\% & 73.9\% & 62.8\%$^{**}$ & 15.4\%$^{**}$ & 1.9\% & 15.0\%$^{*}$ & 0.7\%  & 22.0\%$^{**}$ & 7.8\%  \\ 
  & \textbf{All Removed} & 67.2\% & 76.4\% & 67.9\% & 34.0\% & 9.3\%  & 29.7\% & 2.8\%  & 42.0\% & 11.5\% \\ 
\midrule
\multirow{4}{*}{GPT-4o} 
  & Whitespaces Removed & 63.7\% & 79.5\% & 63.3\% & 14.2\% & 1.1\%  & 8.4\%  & -0.2\%  & 9.4\%  & 4.0\%  \\ 
  & Indentations Removed & 63.7\% & 78.6\% & 63.3\% & 8.2\%  & 0.1\%  & 8.1\%  & -0.4\%  & 12.5\% & 3.7\%  \\ 
  & Newlines Removed & 62.4\% & 79.2\% & 65.4\% & 7.9\%  & 0.1\%  & 6.1\%  & -0.8\%  & 8.4\%  & 3.8\%  \\ 
  & \textbf{All Removed} & 61.8\% & 77.7\% & 65.5\% & 33.7\% & 0.4\%  & 26.2\% & -0.5\%  & 33.9\% & 4.6\%  \\ 
\bottomrule
\end{tabular}
{\footnotesize \makebox[\linewidth][r]{$^{*}$ : $p$-value $< 0.05\quad\quad$ $^{**}$ : $p$-value $< 0.01\quad\quad\quad\quad$}}
\vspace{-4mm}
\label{tab:rq2}
\end{table*}

While the removal of overall formatting had minimal performance impact, the mixed results—ranging from performance drops to improvements—prompted us to figure out the impact of each individual formatting element.
To this end, we conduct an ablation study on the three top-performing commercial LLMs (Claude-3.7, Gemini-1.5, and GPT-4o) from our prior experiments.
The ablation study involves removing one type of formatting element at a time—indentation, whitespaces, or newlines—and observing the resulting performance of the LLMs.
This study focuses on C++, Java, and C\#, excluding Python, as its syntax only allows for whitespace removal, leaving no room for ablation.
Specifically, for each sample in the benchmark, we generate three ablated versions: Whitespaces Removed, Indentations Removed, and Newlines Removed.
Similar to the experiments in RQ1, we feed these versions to the LLMs and compute the Pass@1 and token reductions for both inputs and outputs.
Additionally, we assess the statistical significance between the Pass@1 scores and token counts of the experimental group (one formatting element removed) and the control group (all formatting elements removed). 
The results are presented in~\Cref{tab:rq2}.

\subsubsection{Impacts on LLM performance}
Across the three commercial LLMs evaluated, we observe that the impact of removing individual formatting elements differs across models.
Claude-3.7 and GPT-4o demonstrate remarkable stability in performance when individual formatting elements are removed, where all $p$-values are all higher than 0.05, indicating no significant differences.
Specifically, across all languages, Claude-3.7's performance varies by less than 1\% when whitespaces or indentations are removed and even improves by an average of 1.3\% when only newlines are removed.
Similarly, GPT-4o exhibits minimal performance fluctuations, with an average variation of 0.8\% across all removal strategies, highlighting its robustness to formatting changes.
However, Gemini-1.5 shows greater sensitivity to formatting removal on a single type of formatting element, with some settings, such as whitespaces removed C\# and newlines removed Java, showing statistically significant Pass@1 drops compared to removing all elements.
Compared to fully unformatted code, its average Pass@1 across all three languages declines 4.0\% with whitespace removal, 3.1\% with indentation removal, and 2.9\% with newline removal.
This indicates that Gemini-1.5 is more sensitive to the removal of individual formatting elements, particularly in C\# and Java.
These findings indicate that our prior observations—that code formatting does not negatively impact LLM performance—do not extend to the removal of individual formatting elements.
We hypothesize that partially unformatted code is less common in training datasets, making it more challenging for some LLMs to process effectively.
Interestingly, certain selective removal strategies yield better performance than both fully formatted and unformatted code for Claude-3.7 and GPT-4o.
For instance, when only newlines are removed in C++, Claude-3.7's Pass@1 increases 1.6\% compared to the version with all elements removed.
These results indicate that these LLMs may have specific preferences for certain formatting elements, which could be an interesting area for future research.

\begin{tcolorbox}[size=title]
    {\textbf{Takeaway\#5:}}
In terms of individual formatting elements, removing such an element may introduce a negative impact, as Gemini-1.5 exhibits a significant performance degradation.
\end{tcolorbox}

\subsubsection{Impacts on Token Reduction}
Our analysis reveals that different formatting elements contribute unequally to token consumption, with significant variations across both languages and models.
For input code tokens, newlines contribute most significantly to token count for Claude-3.7 and Gemini-1.5, accounting for an average of 14.6\% for Claude-3.7 and 17.5\% for Gemini-1.5 across languages.
For GPT-4o, however, whitespaces have a higher impact with an average of 10.7\% compared to newlines at 7.5\%. 
Indentations represent the second largest contributor with an average of 7.9\% for Claude-3.7, 8.9\% for Gemini-1.5, and 9.6\% for GPT-4o across all languages.
Such differences are caused by the different tokenizers of each LLM.
The input efficiency gains also vary by programming language. 
For instance, newline removal alone reduces Java tokens by 18.7\% for Claude-3.7 and 22.0\% for Gemini-1.5.
C++ follows with an average reduction of 9.2\% across selective strategies, while C\# shows 7.9\%.
Similar to Takeaway \#3, output token reduction remains minimal across all models and formatting strategies.
On average, output tokens decrease by only 0.4\% for Claude-3.7, 3.8\% for Gemini-1.5, and 1.3\% for GPT-4o when individual formatting elements are removed.

\begin{tcolorbox}[size=title]
    {\textbf{Takeaway\#6:}}
Removing individual elements can also reduce input tokens by a considerable amount, but they still suffer from the same issue in output token efficiency as completely unformatted code.
\end{tcolorbox}
\begin{table*}[t]
\setlength{\tabcolsep}{4pt}
\centering
\caption{Comparison of Pass@1 and token reduction across three settings: unformatted input code with the original instruction (Origin) and two experimental prompts (P1 and P2).}
\begin{tabular}{l l l l l l l l l}
\toprule
\multirow{2}{*}{\textbf{Model}} & \multirow{2}{*}{\textbf{Language}} & \multicolumn{3}{c}{\textbf{Pass@1}} & \multirow{2}{*}{\textbf{Input Reduction}} & \multicolumn{3}{c}{\textbf{Output Reduction}} \\
\cmidrule(lr){3-5} \cmidrule(lr){7-9}
 &  & \textbf{Origin} & \textbf{Prompt(P1)} & \textbf{Prompt(P2)} &  & \textbf{Origin} & \textbf{Prompt(P1)} & \textbf{Prompt(P2)} \\
\midrule
\multirow{4}{*}{\centering Gemini} 
 & C++      & 67.2\%  & 47.4\%$^{**}$  & 11.1\%$^{**}$  & 34.0\%$^{**}$  & 9.3\%   & 25.4\%$^{**}$  & 29.2\%$^{**}$  \\
 & C\#      & 76.4\%  & 48.7\%$^{**}$  & 4.4\%$^{**}$   & 29.7\%$^{**}$  & 2.7\%   & 21.0\%$^{**}$  & 25.1\%$^{**}$   \\
 & Java     & 67.9\%  & 34.6\%$^{**}$  & 14.6\%$^{**}$  & 42.0\%$^{**}$  & 11.5\%  & 37.2\%$^{**}$  & 38.9\%$^{**}$  \\
 & Python   & 71.8\%  & 68.1\%  & 10.9\%$^{**}$  & 5.6\%   & 4.1\%   & 5.5\%   & 9.9\%  \\
\midrule
\multirow{4}{*}{\centering GPT-4o} 
 & C++      & 61.8\%  & 69.7\%$^{**}$  & 65.9\%  & 33.7\%$^{**}$  & 0.4\%   & 1.2\%   & 32.6\%$^{**}$ \\
 & C\#      & 77.7\%  & 85.8\%$^{**}$  & 82.7\%$^{*}$  & 26.2\%$^{**}$  & -0.5\%   & 0.1\%   & 25.6\%$^{**}$ \\
 & Java     & 65.6\%  & 68.5\%$^{*}$  & 66.5\%  & 33.9\%$^{**}$  & 4.6\%   & 5.4\%   & 36.1\%$^{**}$ \\
 & Python   & 71.5\%  & 80.9\%$^{**}$  & 59.7\%$^{**}$  & 9.4\%   & 0.4\%   & 0.4\%   & 14.4\%$^{**}$ \\
\bottomrule
\end{tabular}

{\footnotesize \makebox[\linewidth][r]{$^{*}$ : $p$-value $< 0.05\quad\quad$ $^{**}$ : $p$-value $< 0.01\quad\quad\quad\quad\quad\quad$}}
\label{tab:prompt_tuning_single}
\end{table*}

\subsection{RQ3: Methods for Token-Efficient Generation}

As demonstrated in RQ1 and RQ2, even when the input code is unformatted, LLMs tend to generate formatted code in their outputs, resulting in insufficient code reduction.
To explore whether and how this issue can be mitigated, we experiment with two techniques: prompt engineering and fine-tuning.
The experiments were performed on two SOTA commercial LLMs, GPT-4 and Gemini-1.5, because they are the only commercial models, to the best of our knowledge, offering APIs for fine-tuning.
In the following, we detail the experimental setup for each method and present their respective results.

\subsubsection{Prompting with instructions to generate unformatted code}
Since LLMs can interpret user instructions in prompts, a natural approach is to explicitly request unformatted code outputs.
To explore this direction, we design two distinct instructions: one concise (\textbf{P1:} ``Output code without formatting, maintaining syntax.'') and one detailed and explicit (\textbf{P2:} ``Please directly output the following code, deleting all spacings, newlines, and indentations, provided that it does not violate any syntax rules:'').
Each instruction is appended to the existing FIM task prompts.
Using this revised setup, we evaluate the LLM-generated code completions against unformatted benchmark code in three languages, C++, C\#, and Java, measuring pass@1 performance and quantifying reductions in input and output token counts.
We also compute the statistical significance between the pass@1 scores of the experimental group (using the revised prompt) and the control group (using the original prompt).

As shown in~\Cref{tab:prompt_tuning_single}, prompting can be effective in addressing the insufficient output code token reduction.
GPT-4o, instructed with P2, is observed to successfully achieve substantial output token reduction while maintaining its performance in Java, C++, and C\#.
Specifically, it reduces output code tokens by an average of 27.2\% when processing unformatted code inputs with P2, which is comparable to its input reduction percentage. 
Meanwhile, its performance on these three languages is maintained or even significantly improved.
It demonstrates the feasibility of using prompts for more output token reduction.
However, in other settings, the LLMs cannot work well.
For example, the concise prompt P1 is misunderstood by GPT-4o, resulting in output reductions similar to the original prompt, with the highest reduction being only 5.4\% across all languages.
Gemini-1.5 experiences severe performance degradation under both P1 and P2, despite achieving satisfying output token reductions.
Upon closer inspection, we find that Gemini-1.5 tends to remove elements in a way that violates syntax rules.
For example, 
when given a C\# method declaration like \lstinline!static bool HasCloseElements(List<double> numbers, double threshold)!, Gemini-1.5 generates 
\lstinline!staticboolHasCloseElements(List<double>numbers,doublethreshold)!
where it incorrectly removes the spaces between keywords and types. 
This creates syntax errors as neither \lstinline!staticbool! nor \lstinline!doublethreshold! are valid constructs in C\#.
These failure cases highlight the importance of clear instructions for LLMs.

\begin{tcolorbox}[size=title]
    {\textbf{Takeaway\#7:}}
Prompting LLMs with well-crafted prompts to request unformatted output code can effectively reduce output code tokens while maintaining model's performance.
\end{tcolorbox}

Moreover, appending these instructions as new prompts introduced an overhead in input token count.
Specifically, P1 and P2 added 8 and 28 tokens, respectively, to the input prompt, as measured by GPT-4o's tokenizer.
This creates a break-even point where the method becomes cost-effective only when the reduction in output tokens offsets the additional input cost. 
Notably, output tokens are usually priced higher than input tokens (e.g., 4× higher in GPT-4o) due to the nature of LLMs' inference mechanism.
It implies that the overhead from additional input tokens can be less significant.

However, the length of the prompt not only affects overhead but also determines the space available for clearly describing instructions, presenting an efficiency-performance trade-off.
For example, the short prompt P1, despite having less overhead, was misunderstood by GPT-4o due to its insufficient information.

\begin{tcolorbox}[size=title]
    {\textbf{Takeaway\#8:}}
Prompt design should balance the trade-off between the overhead from prompt length and the clarity of instructions.
\end{tcolorbox}

\subsubsection{Finetuning with unformatted samples}
Fine-tuning is a widely used technique for adjusting model behavior.
In this experiment, we fine-tune the two LLMs, GPT-4o and Gemini-1.5, using unformatted code samples.
We also include a control group for measuring the impact of fine-tuning on model performance, i.e., the same model fine-tuned with formatted code samples in the same fine-tuning environment.
Due to budget constraints, the fine-tuning is limited to Java.
Specifically, from the McEval benchmark, we randomly select 50 Java samples as the training dataset and retain the remaining 305 samples as the test set.
The training dataset is processed into two versions: formatted and unformatted.
For each model, we train two variants, each using one version of the dataset.
The training employs a parameter-efficient fine-tuning method, QLoRA~\cite{dettmers2023qlora}.
We evaluate each fine-tuned model on the retained test set and compare its performance in Pass@1 and token reductions.
We also compute the statistical significance between the pass@1 scores and token counts of the experimental group (fine-tuned with unformatted code samples) and the control group (fine-tuned with formatted code samples).

The results are reported in~\Cref{tab:finetuning_results}.
Similar to prompt engineering, fine-tuning can also significantly increase the output token reduction.
Specifically, Gemini-1.5 achieves a substantial 35.9\% reduction in output tokens with only a minimal performance impact and a 0.4\% insignificant difference in Pass@1,
while GPT-4o reduces output tokens by 24.8\% and even shows a slight performance improvement (2.6\%) when trained on unformatted code.

\begin{tcolorbox}[size=title]
    {\textbf{Takeaway\#9:}}
Fine-tuning with unformatted code can successfully reduce output tokens while maintaining or even improving Pass@1, offering a practical optimization strategy.
\end{tcolorbox}

\begin{table}[t]
\centering
\caption{Comparison between models fine-tuned on formatted code (F) and unformatted code (U).}
\begin{tabular}{lcccc}
    \toprule
    \multirow{2}{*}{\textbf{Model}} & \multicolumn{2}{c}{\textbf{Reduction}} & \multicolumn{2}{c}{\textbf{Pass@1}} \\
    \cmidrule(lr){2-3} \cmidrule(lr){4-5}
    & \textbf{Input} & \textbf{Output} & \textbf{Finetuned(F)} & \textbf{Finetuned(U)}\\ 
    \midrule
    Gemini       & 44.2\%$^{**}$ & 35.9\%$^{**}$ & 64.1\% & 63.7\% \\
    GPT-4o       & 36.5\%$^{**}$ & 24.8\%$^{**}$ & 64.8\% & 67.4\% \\
    \bottomrule
\end{tabular}
{\footnotesize \makebox[\linewidth][r]{$^{*}$ : $p$-value $< 0.05\quad\quad$ $^{**}$ : $p$-value $< 0.01\quad$}}
\vspace{-4mm}
\label{tab:finetuning_results}
\end{table}

\subsubsection{Prompting or Finetuning?}
In RQ3, we experiment with two approaches, prompt engineering and fine-tuning, to reduce more tokens in the output.
The results demonstrate that both methods are viable for achieving this objective.
Each approach has distinct strengths and limitations, which we discuss in detail below.

Prompt engineering guides the LLM in minimizing unnecessary formatting tokens by providing explicit natural language instructions.
This method modifies only the input text without altering the underlying model, making it highly flexible and cost-effective, as it incurs no additional training expenses.
However, our experiments reveal that the effectiveness of prompt engineering depends heavily on the quality of the prompt and the model's capabilities.
Additionally, the prompt itself introduces token overhead, which can limit its practicality during prompt design.

In contrast, fine-tuning the model using unformatted code samples can also optimize token efficiency effectively.
Our experiments show that lightweight fine-tuning with just 50 training samples, combined with PEFT techniques, achieves comparable token reduction.
This approach avoids the overhead associated with prompting.
However, fine-tuning requires access to the model, which may not be feasible for individual users of LLM services.
Furthermore, fine-tuned models tend to exhibit fixed behavioral patterns based on the training data, reducing their adaptability to diverse needs.

In conclusion, the choice between prompt engineering and fine-tuning depends on the LLM's usage scenario and the user's role.
For users seeking quick, flexible solutions without model access or additional training resources, prompt engineering offers a practical and cost-effective option.
On the other hand, fine-tuning provides a more robust and consistent solution for token reduction, particularly for users with the resources and access to modify the model, as well as for domain-specific tasks like code completion.

\begin{tcolorbox}[size=title]
    {\textbf{Takeaway\#10:}}
The choice between prompt engineering and fine-tuning depends on the user's needs and resources.
Prompt engineering is a flexible, cost-effective solution for users without model access, while fine-tuning offers a more robust and consistent approach for those with the resources to modify the model, particularly in stable tasks like code completion.
\end{tcolorbox}

\section{Tool}
Based on the promising results from our experiments, we developed a code transformation tool designed to either remove or restore formatting in source code.
This tool can convert source code into a compact, unformatted version optimized for efficient model comprehension, as well as revert it back to a human-readable format.

\subsection{Implementation}
We implemented a transformation tool supporting C++, Java, C\#, and Python.
It is implemented to serve as an additional pre-processing and post-processing step to minimize the token consumption from the code format.
The tool leverages language-specific formatters: for C-family languages (C++, Java, C\#), we extended \textbf{Uncrustify}~\cite{Uncrustify}, a widely-used formatter.
For Python, we implement a custom formatter built on \textbf{YAPF}~\cite{YAPF}, which accurately handles Python’s indentation-based syntax.
Currently, the tool supports the removal and restoration of three formatting elements: indentation, whitespaces, and newlines.
It can be configured to remove some or all formatting elements to the greatest extent possible without violating syntax rules.

To handle partial code, which is often syntactically incorrect but common in code completions or user prompts, we developed a hybrid solution.
The tool first separates the last unfinished block of code from the main code body and applies different strategies to remove or restore formatting in the split unfinished block and the remaining code.
For the remaining code, we implemented a syntax repair component that uses a bracket-matching mechanism to identify and fix unbalanced brackets.
This helps to correct the syntax in typical cases, such as unbalanced brackets, and enables the code to be processed by our internal formatters.
After processing, the added brackets used for formatting are removed.
For the unfinished block, predefined regular expressions are applied to identify positions where formatting elements can be added or removed.
Finally, the two blocks are concatenated back into a single unit.

\subsection{Usage Scenarios}

Our tool facilitates bidirectional transformation between human-readable, well-formatted code and LLM-friendly, compact code.
As demonstrated in~\Cref{fig:tool_architecture}, our tool can be used to build a dual-conversion inference workflow, enabling LLMs to benefit from the efficient compact code while still allowing human developers to work with familiar code.
In this workflow, our tool removes the formatting elements in the input code, reducing token consumption for LLMs to understand. 
For the output, it restores the formatting in LLM-generated code to improve human readability without altering the underlying logic.

It can be used by both LLM service providers and their users. 
Providers such as OpenAI and Anthropic can leverage this tool to reduce computational overhead on their servers, leading to faster response times, lower resource utilization, and enhanced service efficiency and scalability. 
Furthermore, developers and organizations utilizing LLMs can integrate our tool as an additional pre-processing and post-processing step to minimize unnecessary token consumption.
It can reduce their financial costs since most LLM APIs are charged based on token usage.

\subsection{Performance Testing}
We test our tool using the samples from the McEval dataset.
Specifically, we compare the AST of each code sample before and after the transformation and achieve 100\% AST equivalence across all test cases, confirming that our transformations preserve semantic correctness while only modifying formatting elements.
We chose AST equivalence over text comparison since it guarantees program behavior equivalence, whereas textual comparison might be overly sensitive to non-functional formatting differences.
During the test, our tool achieves an average transformation speed of 76ms per code sample, which is a negligible overhead.
Such efficiency is crucial for real-time applications, like IDE plugins or API middleware, where transformation latency could otherwise impact user experience.

\section{Related Work}
\subsection{Program Simplification}
Prior work on optimizing code representation for LLMs has explored various approaches. 
One common strategy involves training a specialized Byte-Pair Encoding tokenizer\cite{sennrich2016neuralmachinetranslationrare} on a code corpus, which can reduce token count compared to tokenizers trained on natural language corpora\cite{wang-etal-2021-codet5}. However, this approach still suffers from unnecessary formatting tokens, and once a model is trained, its tokenization strategy remains fixed, limiting adaptability.
Some program simplification methods remove parts of input code based on auxiliary models\cite{sivand,p2im} or attention weights\cite{Zhang_2022}. These methods inevitably compromise the semantic integrity of the code and are irreversible, restricting their applicability to code understanding tasks.
SimPy\cite{sun2024aicodersusrethinking} introduces the concept of AI-oriented grammar for compact code representation. While achieving substantial token reduction, this method modifies the syntax of the original programming language, requiring a specialized parser and model retraining to work with the new grammar.
In this work, we propose a plug-and-play method for optimizing LLM token usage in code while preserving complete program semantics. Through empirical experiments, we reveal that SOTA models show resilience to the removal of formatting tokens. Building on this insight, we develop a code transformation tool that enables seamless conversion between human-readable and token-efficient representations, allowing LLMs to benefit from the token efficiency of unformatted code.

\subsection{Coding Style with LLM}
Coding style has been extensively studied in the field of software engineering.
Oman et al.\cite{100348.100385} established taxonomies that influenced code development guidelines and formatting tools. Building on these foundations, machine-learning based tools such as CODEBUFF \cite{2997364.2997383} and STYLE-ANALYZER \cite{markovtsev2019styleanalyzerfixingcodestyle} have been developed to enforce consistent code formatting.
Mi et al.\cite{2972958.2972963} measured style inconsistencies within software project teams using clustering methods, while Wang et al. \cite{wang2024functionalcorrectnessinvestigatingcoding} examined style inconsistencies between LLM-generated code and human-written code.
These researches focused on improving the style consistency of code.
Beyond this, Hu et al.\cite{3691620.3695072} investigated the impact of poor readability on LLMs, demonstrating that obfuscation techniques, such as modifying identifier names and injecting dead branches, can degrade model performance. However, their work did not explore the role of formatting elements.
Unlike previous research, our work offers a unique perspective by quantifying the impact of formatting elements on computational cost and model performance across multiple languages and architectures. In addition, we analyze the contribution of individual formatting components, offering a deeper understanding of their influence on LLM processing.

\section{THREATS TO VALIDITY}
\subsection{Generalization}
Due to budget and hardware limitations, our experiments were restricted to four programming languages and focused solely on the Fill-in-the-Middle (FIM) task.
As a result, our findings may not generalize to languages with distinct formatting styles or other LLM application scenarios.
However, the selected languages are widely adopted in practice, and the FIM task is a common benchmark for nearly all coding assistants in IDEs.
These choices provide a reasonable foundation for evaluating the model's performance in real-world settings, where enhancing token efficiency can lead to significant resource savings and is often a high priority.

\subsection{Non-transparent Commercial LLMs}
Our experiments involved invoking and fine-tuning commercial LLMs, such as GPT-4 and Gemini-1.5, through their closed-source APIs.
This reliance on proprietary systems introduces potential limitations, as the internal mechanisms, training data, and fine-tuning strategies of these models are not transparent.
Consequently, reproducibility and detailed analysis of their behavior are challenging, which may affect the generalizability of our findings.
Future work could explore open-source alternatives or collaborate with API providers to gain deeper insights.

\section{Conclusion and Future Work}

This paper exposes the hidden costs of code readability in LLM processing, demonstrating that formatting elements consume approximately 24.5\% tokens across languages while providing minimal benefits for advanced models.
Our analysis identifies the contributions of three kinds of formatting elements, whitespace, indentation, and newlines, and shows that both fine-tuning and prompting on unformatted code can further reduce token usage without compromising quality.
These findings challenge conventional views of code formatting as purely human-oriented and reveal opportunities for substantial efficiency improvements in LLM-powered development workflows.
Our bidirectional transformation tool offers a practical solution for balancing human readability with computational efficiency.
In the future, we will investigate how formatting impacts more complex reasoning tasks beyond code completion.

\begin{acks}
This research / project is supported by Xiaoning Du’s Google Research Scholar Program Award and the National Research Foundation, under its Investigatorship Grant (NRF-NRFI08-2022-0002).
Any opinions, findings and conclusions or recommendations expressed in this material are those of the author(s) and do not reflect the views of National Research Foundation, Singapore.
\end{acks}

\balance
\bibliographystyle{ACM-Reference-Format}
\bibliography{references}

\end{document}